\documentclass{aa} 
\usepackage{color}
\usepackage{graphicx}
\usepackage[]{natbib}
\usepackage{sidecap}
\usepackage{txfonts}
\begin{document}
\title{Flux emergence and coronal eruption}

\author{V. Archontis
        \inst{1}
        \and
        A. W. Hood\inst{2}}

 \institute{School of Mathematics and Statistics, University of St. Andrews, North Haugh, 
            St. Andrews, Fife, KY16 9SS, UK\\
            \email{vasilis@mcs.st-and.ac.uk}
            \and
            \email{alan@mcs.st-and.ac.uk}}

\date{Received ...; accepted ...}

\abstract
{}
{Our aim is to study the photospheric flux distribution of a twisted flux tube that emerges from the solar interior. We also report on the 
eruption of a new flux rope when the emerging tube rises into a pre-existing magnetic field in the corona.} 
{To study the evolution, we use 3D numerical simulations by solving the time-dependent and resistive MHD equations.
We qualitatively compare our numerical 
results with MDI magnetograms of emerging flux at the solar surface.}
{We find that the photospheric magnetic flux distribution consists of two regions of opposite polarities and 
elongated magnetic tails on the two sides of the polarity inversion line (PIL), depending on the 
azimuthal nature of the emerging field lines and the initial field strength of the rising tube. 
Their shape is progressively deformed due to plasma motions towards the PIL. Our results are in qualitative agreement 
with observational studies of magnetic flux emergence in active regions (ARs). Moreover,  
if the initial twist of the emerging tube is small, the photospheric magnetic field develops an undulating shape and does not possess
tails. In all cases, we find that a new flux rope is formed above the original axis of the emerging tube that may erupt 
into the corona, depending on the strength of the ambient field.}
{}

\maketitle

\section {Introduction}
\label{sec:intro}

Active regions are often associated with episodes of magnetic flux emergence from the solar interior (\citet{zwaan85} and references therein).
An important question then is, what is the evolution of the magnetic field configuration at the photosphere during emergence? 
Observations of emerging flux regions (EFRs) as recorded at the photospheric level, show that they consist 
of two main flux bundles of opposite magnetic polarity that may be the 
manifestation of an emerging flux tube. There is strong 
evidence that, in many EFRs, the rising magnetic fields are twisted. 
The idea of a flux rope configuration has been supported by photospheric measurements and observations
of emerging fields in {\it normal} and {\it complex} 
(the so-called $\delta$ sunspot) ARs \citep{tanaka91, lites95, leka96, can09}.\\
\begin{figure*}[t]
\centering
\includegraphics[width=5.7cm,height=4.cm]{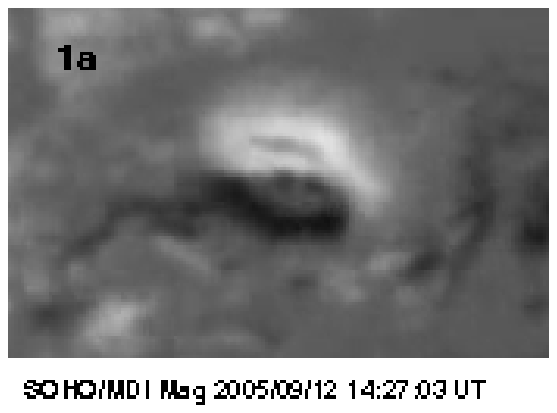}
\hskip 0.01\linewidth
\includegraphics[width=5.7cm,height=4cm]{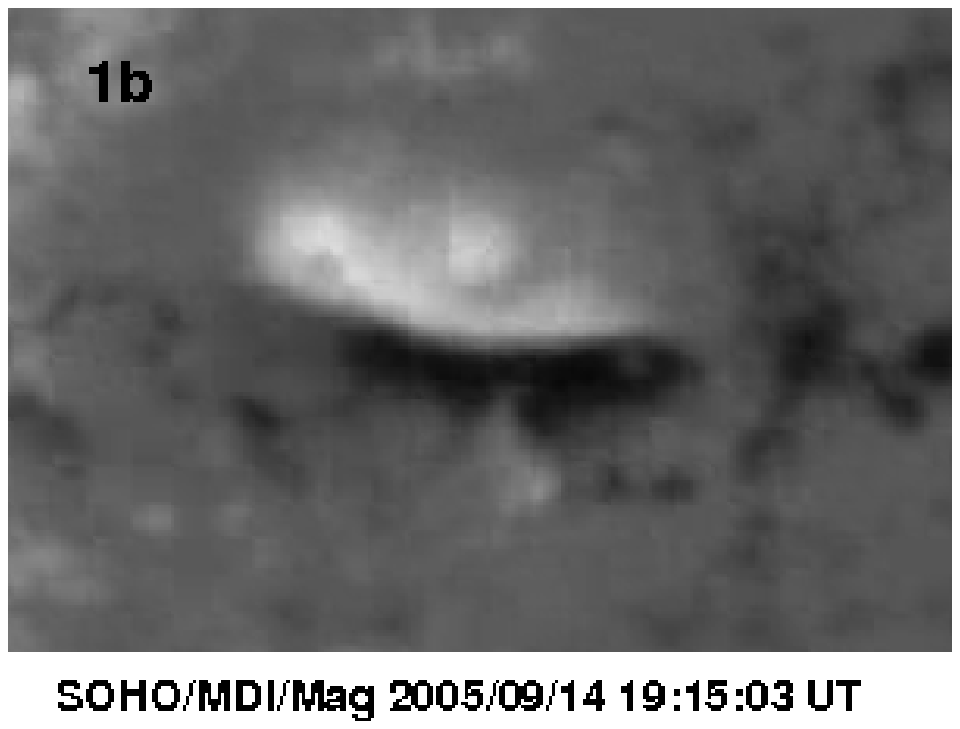}
\hskip 0.01\linewidth
\includegraphics[width=5.7cm,height=4.cm]{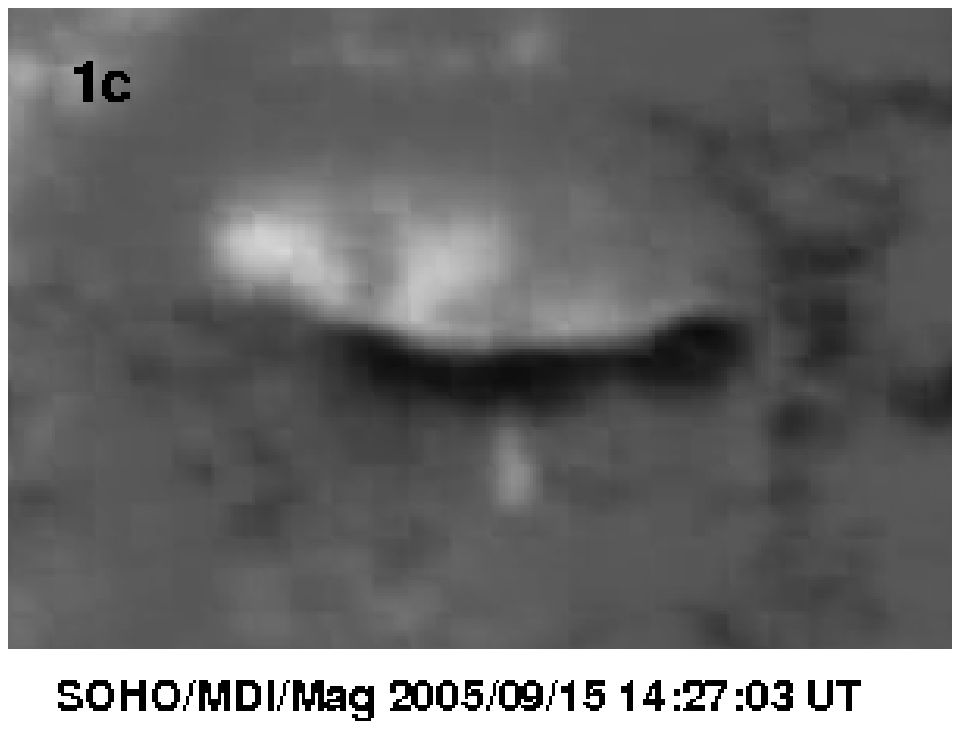}
\hfill
\includegraphics[width=5.7cm,height=3.8cm]{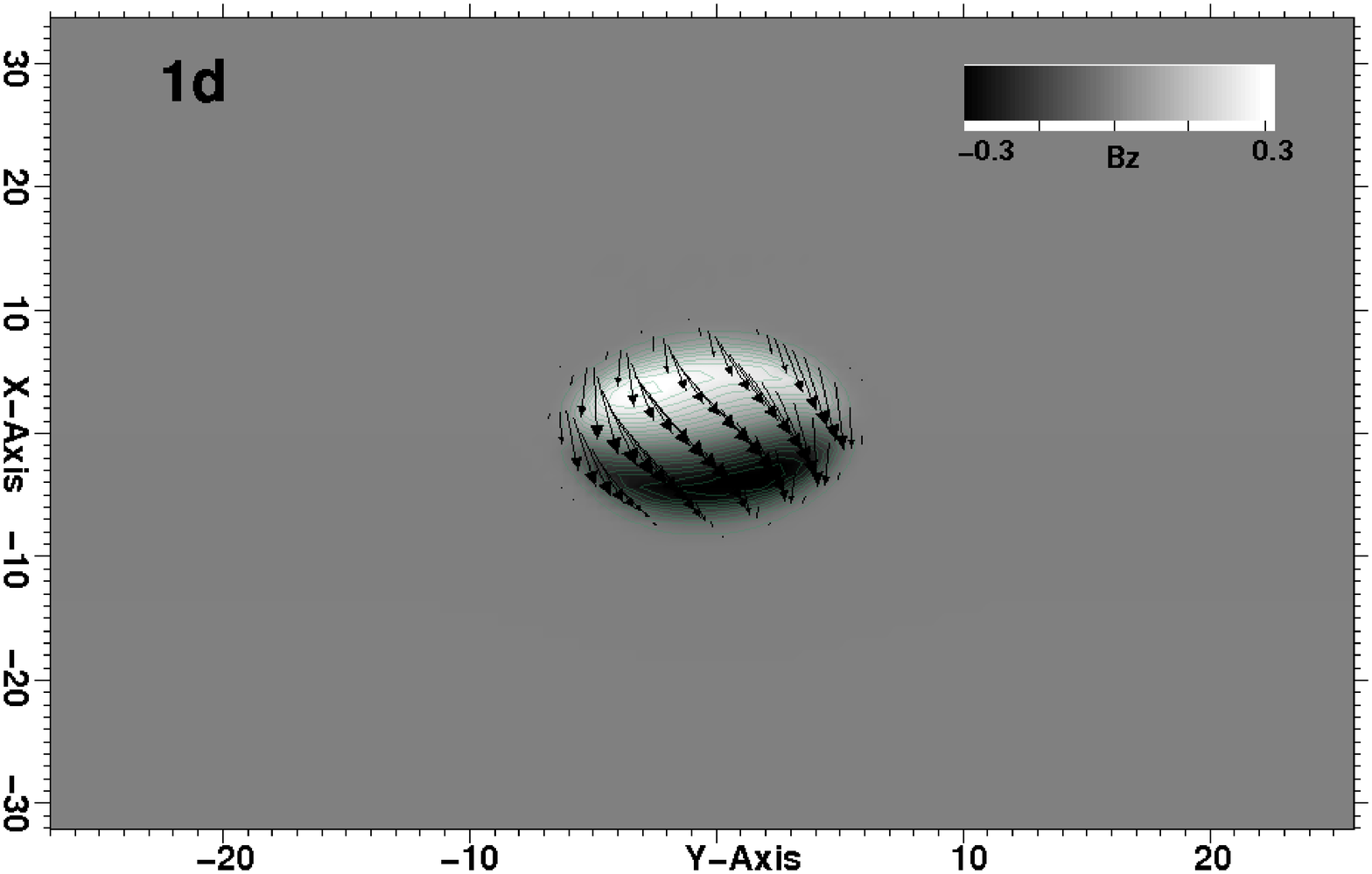}
\hskip 0.01\linewidth
\includegraphics[width=5.7cm,height=3.8cm]{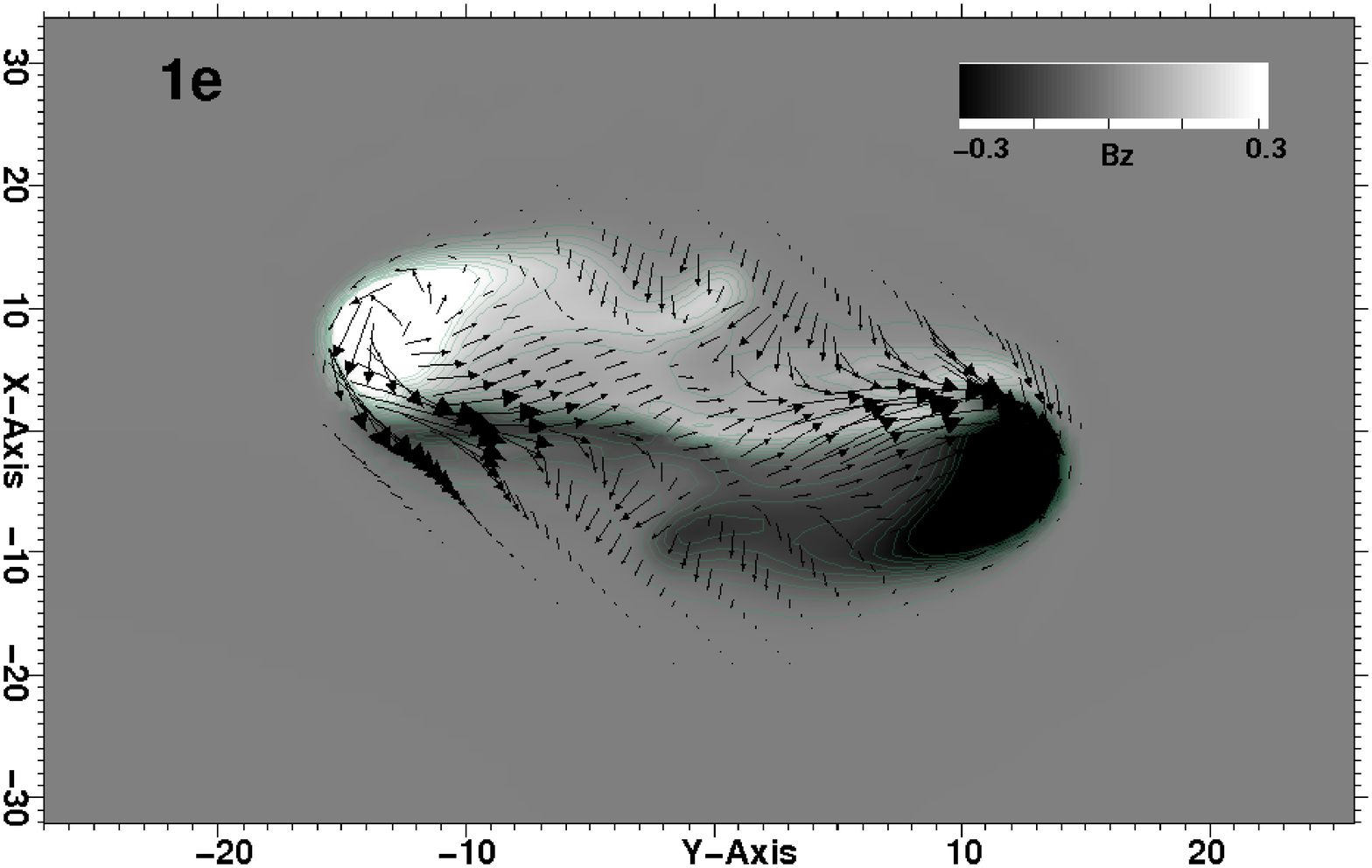}
\hskip 0.01\linewidth
\includegraphics[width=5.7cm,height=3.8cm]{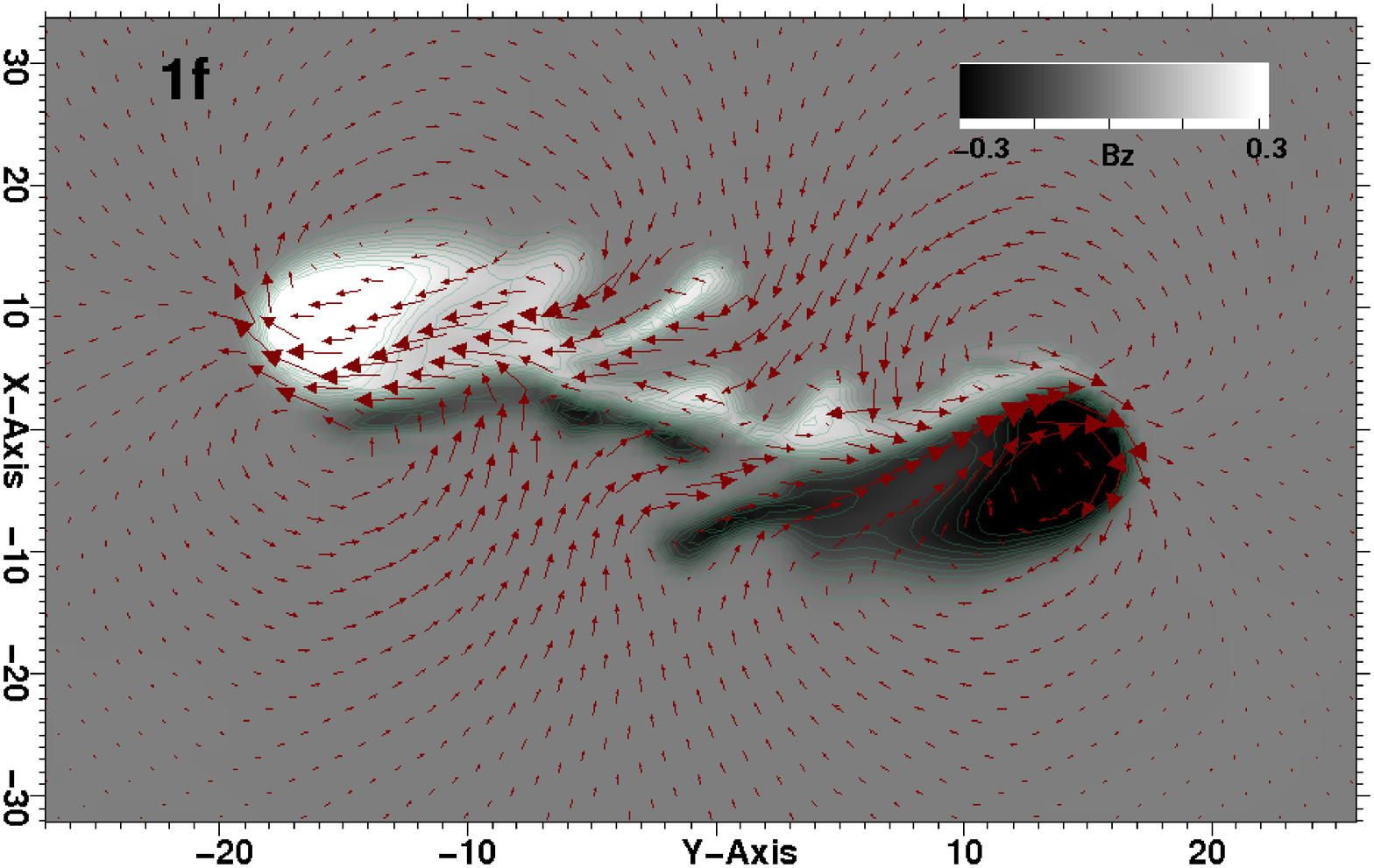}
\caption{{\bf Top:} SOHO/MDI magnetograms during flux emergence in NOAA AR 10808. {\bf Bottom}: {\it Magnetograms}
produced in the numerical experiments, at $z=21$ and $t=40$, $t=125$ and $t=175$ for the panels 1d-1f respectively. Arrows
show the horizontal component of the magnetic field, $B_{hor}$ (panels 1d and 1e, black) and velocity field, $V_{hor}$ (panel 1f, red).}
\end{figure*}
A common feature in EFRs is the presence of magnetic {\it tongues} or {\it tails}, which are connected with the main polarities 
on the two sides of the PIL of the AR \citep{li07, can09, chandra09}. The appearance of magnetic {\it tails} is 
interpreted as the result of the emergence of twisted magnetic field lines at the photosphere \citep{lopez00}. However, a study of how 
the formation and evolution of the {\it tails} depend on the physical properties of the emerging field is still missing.
\citet{can09} used SOHO/MDI magnetograms and they reported on the existence of {\it tails}, which formed along the PIL and accompanied 
the emergence of magnetic flux in the region NOAA AR 1808. The shape of the {\it tails} was deformed during the evolution of the system.
They also used the THEMIS vector magnetogram to reconstruct the coronal field (via a nonlinear force-free model) 
and found evidence for a pre-eruptive twisted flux tube above the emerging field.\\
In this paper, for the first time, we focus on the photospheric distribution of an emerging flux tube and the formation of the 
{\it tails}, showing the relationship between the topology of the {\it tails} and the initial tube parameters. We compare some of the numerical 
results with the observations by \citet{can09} and we find a preliminary, qualitative agreement. 
Secondly, we report on the emergence of the tube into a magnetized corona and the subsequent coronal eruption of a flux rope. 
Similar to previous experiments \citep{mag01a,man04,arc08a,arc08b,hood09}, we find 
that the emerging twisted flux tube and the coronal rope are two distinct structures. More importantly, we find that the evolution of the
erupting rope (ejective vs confined eruption) depends on the strength of the ambient field.
\section{Model}
\label{sec:Model}
The results in our experiments are obtained from a 3D MHD simulation. 
The basic setup of the experiment follows the simulation by \citet{arc05} and consists of a hydrostatic atmosphere and a 
horizontal twisted magnetic flux tube.
All variables are made dimensionless by choosing photospheric values of density,  
$\rho_{\mathrm{ph}}=3\times10^{-7}\,\mathrm{g}\,\mathrm{cm}^{-3}$, pressure,
$p_{\mathrm{ph}}=1.4\times10^{5}\,\mathrm{ergs}\,\mathrm{cm}^{-3}$, and pressure scale height,
$H_{\mathrm{ph}}=170\,\mathrm{km}$, and by derived units (e.g., magnetic field strength
$B_\mathrm{ph}=1300\,\mathrm{G}$, velocity $V_\mathrm{ph}=6.8\,\mathrm{km}\,\mathrm{s}^{-1}$ and
time $t_{\mathrm{ph}}=25\,\mathrm{s}$).
The atmosphere includes a subsurface layer $(-10\le z<20)$, 
photosphere $(20\le z<30)$, transition region $(30\le z<40)$, and corona $(40\le z\le 110)$. The numerical domain 
has a dimensionless size of $[-70,70] \times [-80,80] \times[-10,110]$ in the longitudinal ($x$), transverse 
($y$) and vertical ($z$) directions, respectively. The magnetic flux tube is imposed $1.4 Mm$ below the 
surface along the $y$-axis. The radius of the tube is $425 Km$. The axial field is defined by 
$B_y=B_0\,\mathrm{exp}(-r^2/R^2)$ and $B_\phi=\alpha\,r\,B_y$, where $r$ is the radial distance
from the tube axis and $\alpha$ is the twist per unit length. The twist of the fieldlines around 
the axis of the tube is uniform. The tube is made buoyant by applying
a density perturbation $\Delta\,\rho=[p_t(r)/p(z)]\,\rho(z)\,\mathrm{exp}\,(-y^2/\lambda^2)$, where
$p_t$ is the pressure within the flux tube and $\lambda$ is the buoyant part of the tube. 
We perform four experiments: E1 ($B=3,\alpha=0.4,\lambda=10$), E2 ($B=5,\alpha=0.4,\lambda=10$), 
E3 ($B=5,\alpha=0.4,\lambda=20$) and E4 ($B=5,\alpha=0.1,\lambda=10$).
In the numerical domain, we use periodic boundary conditions in horizontal directions
and closed boundaries with a wave damping layer in vertical directions.

\section {Results}
\subsection {Emergence into the photosphere}
\label{sec:photosphere}

\begin{figure*}[t]
\centering
\includegraphics[width=6.cm,height=5.2cm]{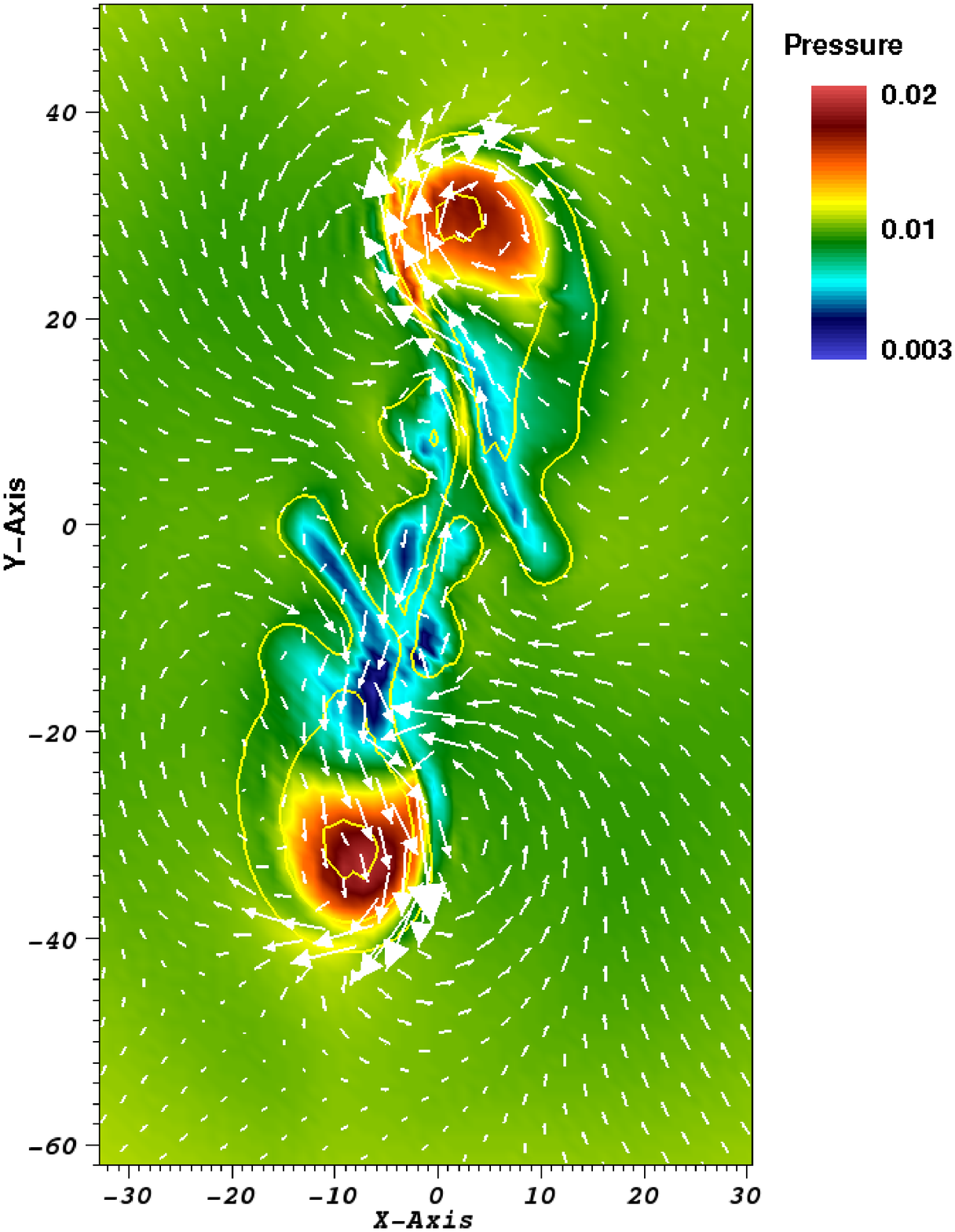}
\hskip 0.01\linewidth
\includegraphics[width=5.cm,height=5.2cm]{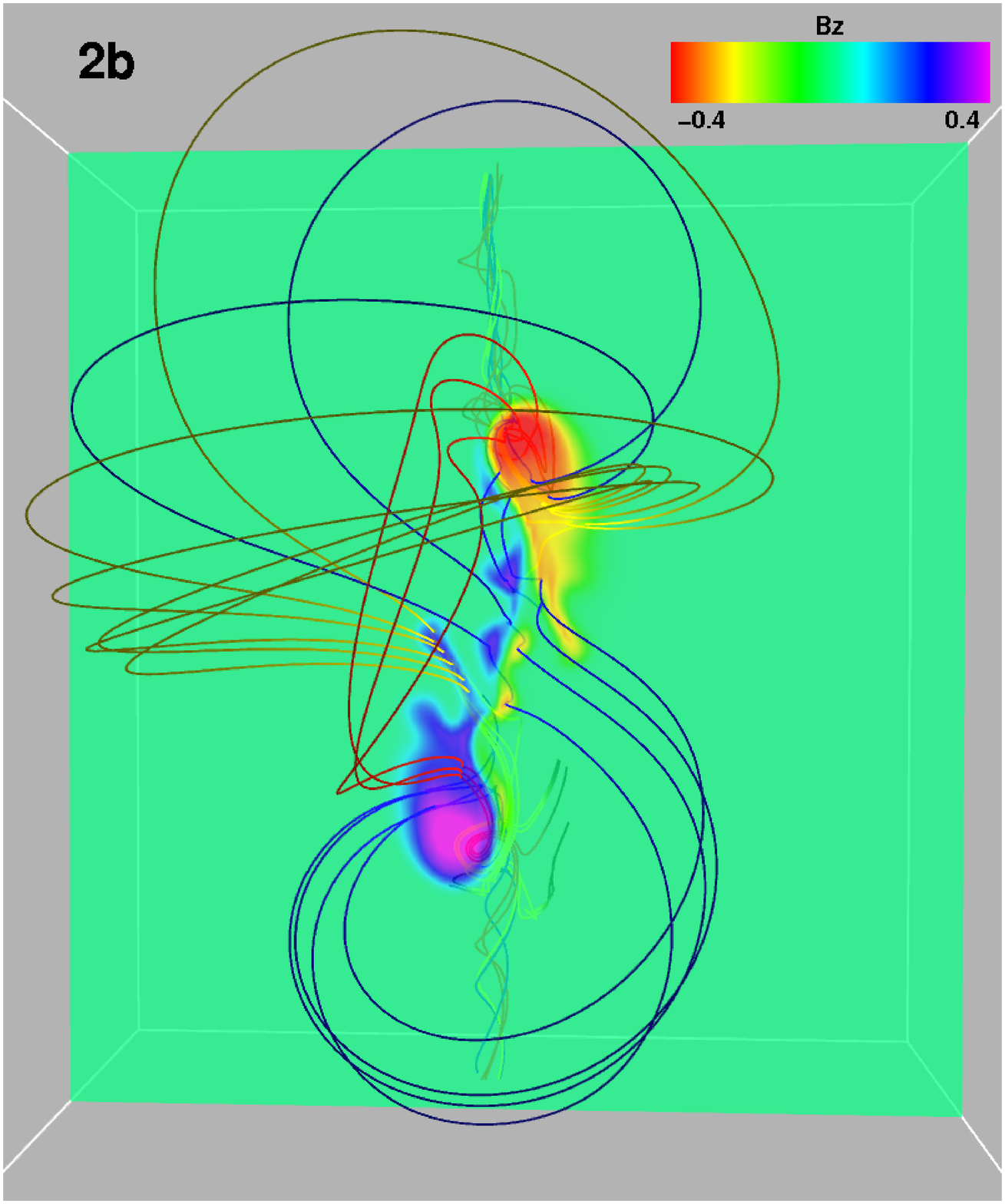}
\hskip 0.01\linewidth
\includegraphics[width=5.5cm,height=5.2cm]{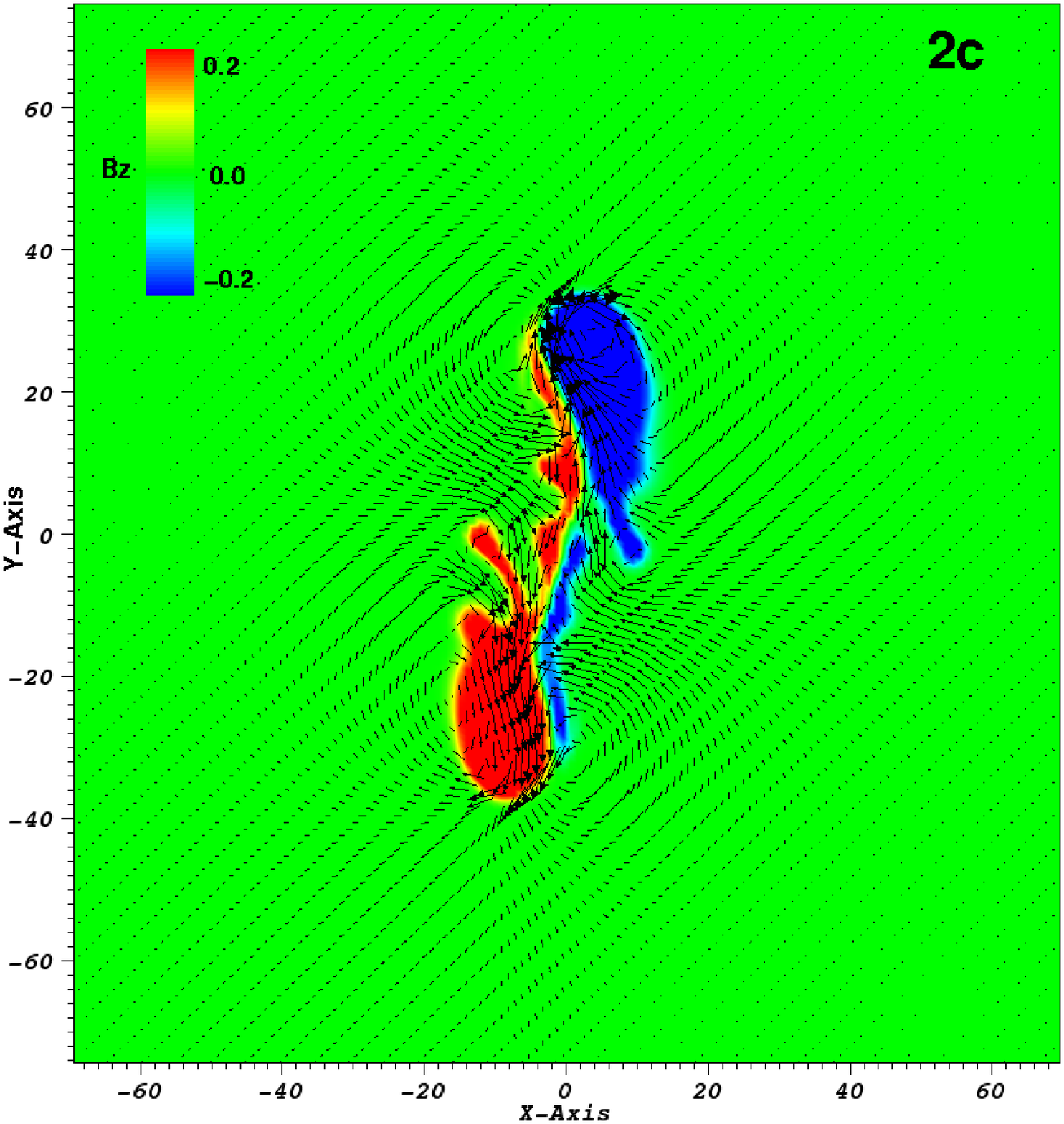}
\hfill
\includegraphics[width=5.5cm,height=5.2cm]{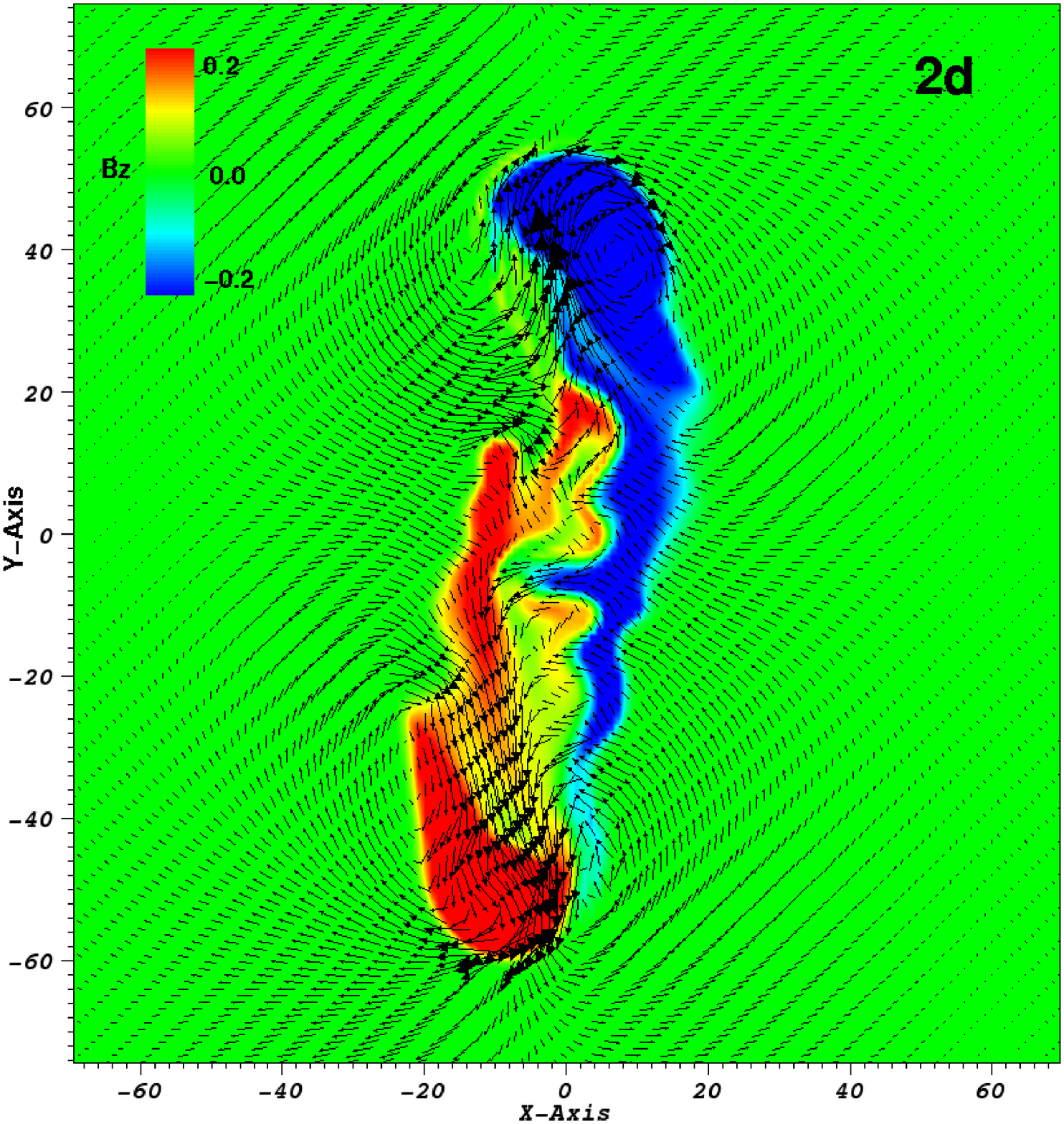}
\hskip 0.01\linewidth
\includegraphics[width=5.5cm,height=5.2cm]{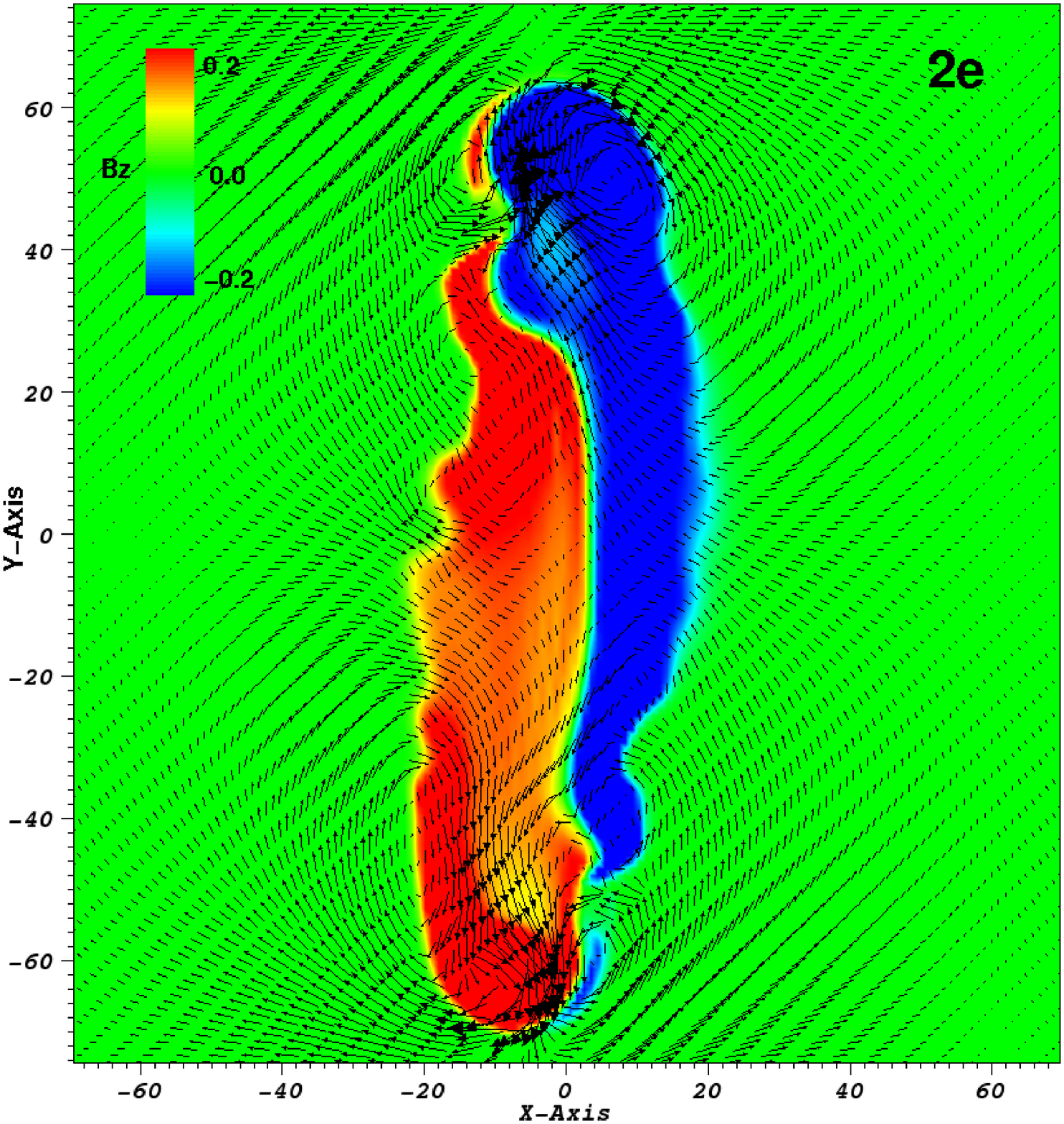}
\hskip 0.01\linewidth
\includegraphics[width=5.5cm,height=5.2cm]{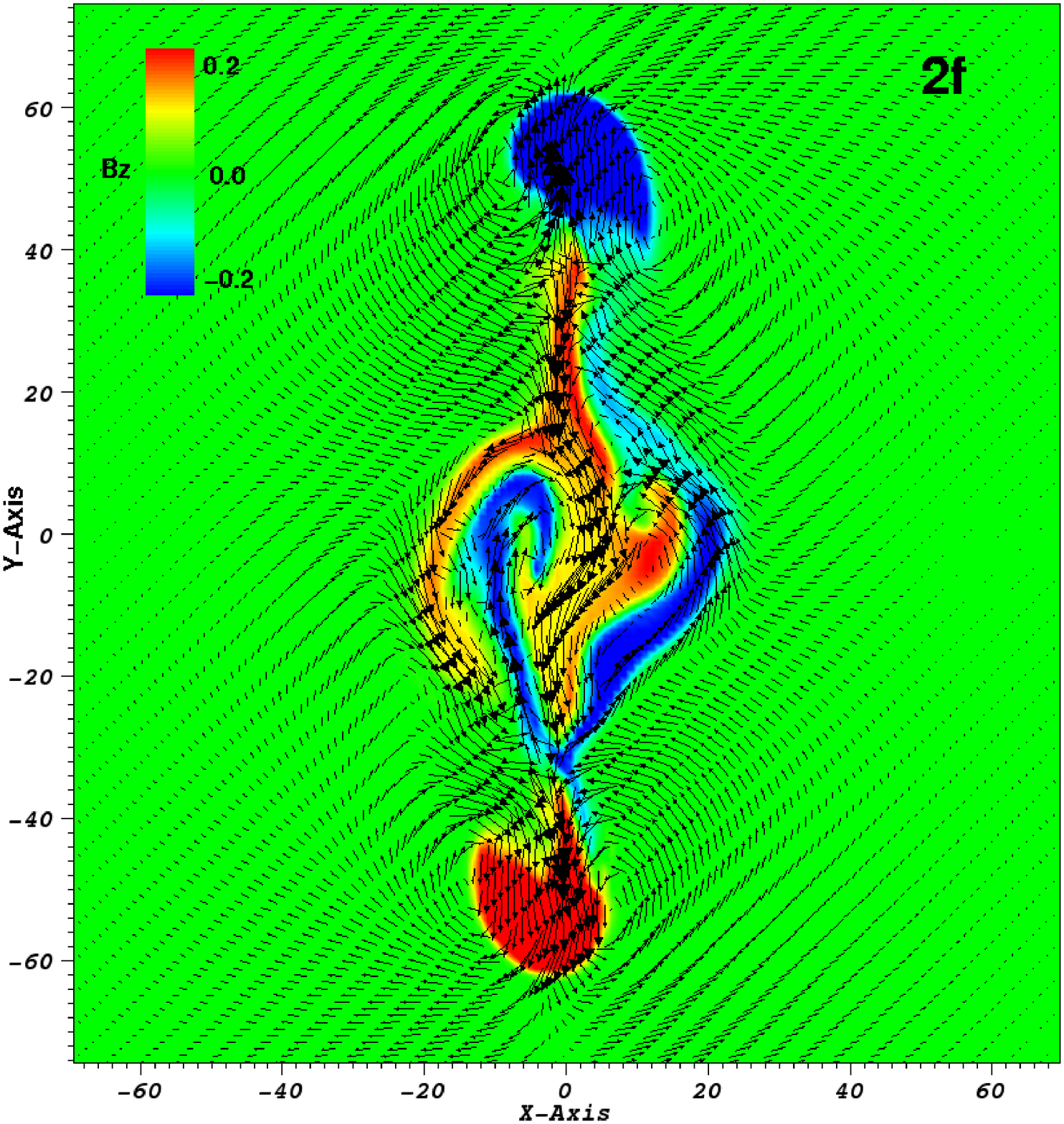}
\caption{{\bf Top:} The colour-scaled maps correspond to the $P_{tot}$ (2a) and $B_{z}$ (2b, 2c). Contours show $B_{z}$ and arrows the horizontal velocity. 
Time is $t=165$ and $z=25$, all for experiment E1. {\bf Bottom:} Distribution of $B_{z}$ for E2 (2d), E3 (2e) and E4 (2f).}
\end{figure*}

Figure 1 (panels 1a-1c) shows the evolution of the emerging field in the NOAA AR 10808. 
At the beginning (panel 1a) there is a clear appearance of a bipolar region at the photosphere with a North-South 
orientation. The two polarities progressively diverge from each other in an approximate East-West direction (panels 1b, 1c).
During the evolution of the system, two elongated {\it tails} or {\it tongues} are formed in the wake of the 
two polarities (panel 1b). Initially, the {\it tails} possess an apparently coherent shape but as time goes on their structure appears 
to be more fragmented on the two sides of the PIL (panel 1c).

Panels 1d-1f show the photospheric distribution of the emerging field in our numerical experiments.
Panel 1d shows the bipolar appearance of the emerging field, shortly after it intersects the photosphere.
The North-South orientation of the bipolar field is due to the strong initial twist of the flux tube. 
Eventually (panel 1e), the two main polarities drift apart toward an East-West orientation. Similar to the observations, they are followed 
by magnetic {\it tails} that develop an intricate geometrical shape. 
The projection of the horizontal component of the magnetic field (arrows) is overplotted onto the {\it magnetograms} 
of panels 1d and 1e. At $t=40$, the direction of the horizontal magnetic field vectors shows a normal configuration, i.e. from the
positive to the negative polarity, at the PIL. Later on, as more magnetic flux emerges from the solar interior, the direction 
of the magnetic field reveals a dominant inverse configuration along the PIL. This is due to the rise of the original axis of the twisted flux tube 
above this height ($z=21$). However, the main axis does {\it not} emerge above $2-3$ pressure scale heights, as has been shown in previous
experiments of flux emergence \citep{fan01, mag01a, arc05}.

At a later stage of the evolution (Panel 1f), the elongated {\it tails} develop fingers seperated by dips along the curved PIL.
In the fingers, the magnetic field remains strong (around $70\%$  of the maximum value of $B_{z}$ at this height). At the
dips, the magnetic field is weak and the plasma density is relatively small. In fact, we find that there is a good correlation
between the location of the dips and sites where plasma is moving in the transverse direction. There, the converging flows
may reach values up to $3 Km/s$ and the kinetic energy density becomes larger than the magnetic energy of the field. Thus, it seems that
the shape of the magnetic {\it tails} is deformed due to inflows that are able to compress and advect the magnetic field.
The origin of the inflows depends on the evolution of the total pressure ($P_{tot}$ = magnetic + gas pressure) at photospheric heights.
Panel 2a shows the distribution of $P_{tot}$ at $z=25$, when the outer magnetic field has expanded into the corona. Due to the rapid expansion,
a total pressure deficit has developed at the central area of the EFR and so the plasma moves towards the small
pressure, and deforms the {\it tails}.

The link between the appearance of the {\it tails} and the topology of the fieldlines is shown in panel 2b. The yellow fieldlines have been traced
from a far edge of the fingers (at $x=-11$, $y=0$).
These are the outermost fieldlines with a strongly azimuthal nature. The
blue fieldlines are highly twisted and are traced from the fingers of the {\it tails} that are closer to the PIL. They make a full turn
around the main axis of the emerging tube connecting the central area of the two tails. The red fieldlines have been traced from the region closer to the
main positive polarity of the field. They are very weakly twisted, possessing an arch-like bundle of fieldlines, joining the two sunspots.
These fieldlines do not go through the {\it tails}. The above configuration shows that the appearance of the {\it tails} is due to the
projection of the azimuthal component of the magnetic field at the photosphere.

Panels 2c-2f show the magnetic flux distribution at the photosphere for the experiments E1-E4 respectively. We take as a reference case the 
E1 and we examine the effect of varying the initial field strength ${\bf B}$, $\lambda$ and $\alpha$ on the appearance of the {\it tails}.
For comparison, we consider the configuration of the field at a certain time for all experiments. The increase of ${\bf B}$ (in E2) results in 
keeping a coherent shape of the {\it tails} for a longer time period: at $t=165$, the {\it tails} in E2 are less fragmented than in E1. This is 
due to the fact that the total pressure within the EFR is large enough for the {\it tails} to be distinctively deformed by the inflows. However, we should 
emphasize that the shape of the {\it tails} is altered at a later time, when the two sunspots have seperated enough and the magnetic field in the 
EFR becomes weak. The increase of ${\lambda}$ (panel 2e) affects the downward tension of the fieldlines upon the buoyant part of the 
emerging field. The tension is less in the E3 and the field is emerging at the photosphere relatively faster. Thus, at a certain time, the 
magnetic field at the photosphere appears stronger in E3 than in E2. As we mentioned above, the stronger the magnetic field the less effective 
is the deformation of the {\it tails'} shape. This is clearly shown in panel 2e, compared to the 
E2 (panel 2d). Also, the appearance of the {\it tails} critically depends on the initial twist of the emerging field. 
In E4, the twist parameter $\alpha$ is equal to 0.1 and the emerging field is almost horizontal and parallel to the E-W direction, 
shortly after its arrival to the photosphere. We find that there is no {\it tail} formation when the emerging field has $\alpha < 0.2$. 
In this case, the EFR consists of the two sunspots and patches of magnetic flux with mixed polarity on the two sides of the PIL. Some of these 
photospheric flux segments are connected with the same fieldlines, possessing an overall undulating magnetic system. This is reminiscent of the 
``sea-serpent'' configuration, which is produced during the emergence of a magnetic flux sheet. The latter develops undulations when it becomes 
unstable to the Parker instability \citep{arc09b}.

\subsection {Eruption into the corona}
\label{sec:corona}

In Section 3.1, we showed that the photospheric fingerprints of the EFR in E1 consist of features (e.g. {\it tails})  
with a similar configuration to observed ARs (e.g. the AR 10808). In addition, the activity in the region NOAA AR 10808 is known to lead 
to filament and CME eruption \citep{can09}. Thus, an important question is whether our twisted flux tube model can produce a
coronal eruption. Our experiment shows that a new flux rope is formed above the original axis of the 
emerging flux tube due to reconnection of sheared fieldlines. The reconnection occurs in the higher photosphere/lower chromosphere in a similar
manner to the model by \citet{van89}. A key issue is whether this eruption is {\it confined} (and, thus, the flux rope cannot fully escape into the 
outer atmosphere) or {\it ejective}. In previous experiments, \citet{arc08b} found that the inclusion of a pre-existing magnetic field in the 
corona may induce a runaway situation, via reconnection, during which the new flux rope fully erupts into the outer solar atmosphere. 
Here, we perform a similar experiment but using different initial parameters for the pre-existing coronal field.
Our aim is to study whether the field strength of the ambient field affects 
the rising motion of the erupting flux rope.

\begin{figure}[htb]
\centering
\includegraphics[width=9cm,height=7cm]{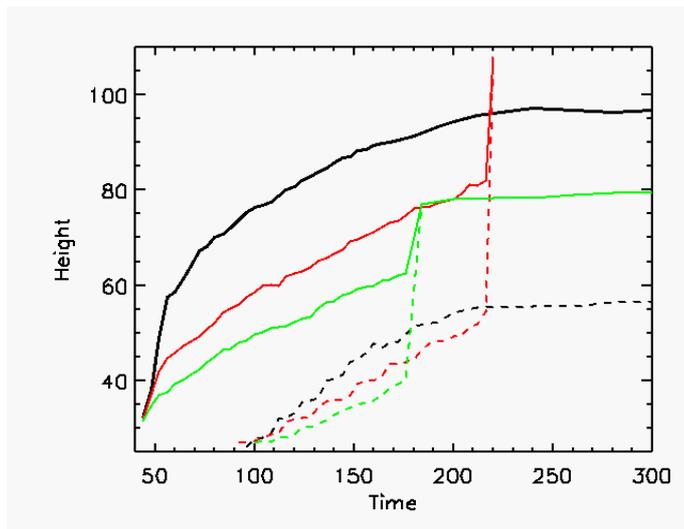}
\caption{Height-time profiles of the apex of the emerging field (solid) and the flux rope (dashed) in experiments 
B1 (black), B2 (red) and B3 (green).}
\end{figure}

The observed magnetogram in the AR 10808 shows that the emerging flux is rising into a pre-existing field oriented in the 
E-W direction. The emerging field has a N-S orientation and, thus, the relative orientation of the two fields is about 90 degrees. 
To simulate this, we include a horizontal and uniform magnetic field in the corona along the y-axis, parallel to the main axis 
of the twisted tube (for example, see \citet{arc05}).
To a first approximation, this field may correspond to the upper part of the observed AR's field, which is likely to be anchored in the 
surrounded diffuse polarities. 
We find that the field strength of the ambient field ($B_{amb}$) plays a critical role in the eruptive motion of the new flux rope. Figure 3 shows the height-time 
profile of the front of the emerging field (solid lines) and the center of the new flux rope (dashed lines) for three experiments (B1, B2 and B3) where: 
$B_{amb}=0.003$ (B1, black lines) 
, $B_{amb}=0.015$ (B2, red lines) and $B_{amb}=0.03$ (B3, green lines) respectively. 
The heights are calculated at the vertical midplane after the emerging field enters the transition region. 

In B1, the front of the expanding tube rises slowly within the magnetized corona and eventually it saturates at a height of $z\approx 96$. 
The new flux rope is formed at the low atmosphere at $t\approx 95$ and, thereafter, it follows a similar evolution to the envelope field of the expanding 
tube. Firstly, it rises 
almost linearly with time but then it reaches an equilibrium where the magnetic pressure force is balanced by the tension of the fieldlines. In this case, 
the eruption is confined: the flux rope is trapped within the envelope field. In B2, the apex of the emerging field reaches 
lower heights during its rising motion. This is because it comes into contact with an ambient field that is stronger and able to delay the emergence. 
At the same time, a considerable amount of the rising magnetic flux is removed from the envelope field due to reconnection. As a result, the distance between 
the new flux rope and the front of the envelope field is reduced. As more magnetic layers above the flux rope are peeled off, the downward tension of the 
envelope fieldlines decreases. Eventually, the flux rope experiences an ejective eruption reaching the upper boundary of the domain very quickly. 
Due to the short distance between the erupting rope and the closed top boundary, the velocity of the center of the flux rope is restricted to $197 Km/s$. 
However, the plasma underneath the flux rope is rising with even higher velocity at $\approx 350 Km/sec$. This is a reconnection jet that is formed due to 
internal (i.e. within the EFR) reconnection of fieldlines and helps the flux rope to accelerate during its eruption. According to these calculations, it is 
possible that the rise of the flux rope might account for a CME-like eruption. 

In B3, the eruption of the flux rope is triggered earlier. Again, this is because the stronger ambient field reconnects
more effectively with the flux above the rope and removes more magnetic layers from the emerging system. However, for the same reason, the front of the envelope 
field rises with a slower rate and the distance between the new flux rope and the front decreases. As a result, soon after the triggering of the 
ejective eruption, the erupting rope collides with the front and loses its distinct circular shape, possibly due to reconnection with the 
ambient field. After the collision, the leading edge of the emerging system is lifted up for a few pressure scale heights. However, it does not reconnect 
effectively with the magnetic flux above it, and eventually reaches a quasi-static state at a height of $z\approx 80$.
Thus, in B3, the ejective flux rope is trapped by the dominant ambient field and not by the envelope field.
\section {Summary and Discussion}
\label{sec:disc}

In this paper, we have presented a 3D model to study the emergence of a twisted flux tube 
throughout the solar atmosphere. Our model gives new insights into the photospheric distribution of the 
emerging magnetic field: it consists of a bipolar region and {\it tails} on the two sides of the PIL. 
The appearance of {\it tails} reveal that the emerging magnetic field is twisted. For small twist, the 
emerging field possess undulations. Our results {\it predict} that the irregular structure of the {\it tails} 
is due to the interplay between the flows and the dynamical evolution of the magnetic field. The configuration of the 
emerging field at the photosphere is in qualitative agreement with observations \citep{can09}.

In agreement with previous simulations, our experiments show the eruption of a flux rope, which is formed above the 
original axis of the emerging tube. For the first time, we find that the field strength of a pre-existing coronal 
magnetic field is a crucial parameter affecting the eruptive phase of the rope. Under the specific conditions of the 
present experiments, we found that the eruption is ejective when $ 0.01<B_{amb}<0.02$. For other values, the 
eruption is confined within the envelope field. 

The aim of these experiments is {\it not} a direct comparison with the observations, but rather to suggest 
possible mechanisms that drive the dynamical behaviour of the system.
Further experiments are required to verify the effect of the initial parameters (e.g. field strength, 
radius, initial atmospheric height and twist, etc.) of the twisted flux tube and the pre-existing field on (a) the characteristics of its photospheric appearance 
(formation and evolution of the {\it tails}, shear and transverse flows, etc.) and (b) the dynamics of the associated eruption.

\begin{acknowledgements}
Financial support by the European Comission through the SOLAIRE network 
(MTRM-CT-2006-035484) is gratefully acknowledged. Simulations were performed on the UKMHD consortium cluster, funded by STFC and a SRIF grant to 
the University of St Andrews.
\end{acknowledgements}

\bibliographystyle{aa}

\end{document}